\documentclass{aastex631}

\newcommand{\B}[1]{{\color{blue}{#1}}}
\definecolor{ForrestGreen}{rgb}{0.133,0.545,0.133}
\newcommand{\G}[1]{{\color{ForrestGreen}#1}}

\usepackage{CJK}

\graphicspath{{./}{figures/}}

\begin{document}
%\begin{CJK*}{UTF8}{gbsn}
\title{Simultaneous Eruption and Shrinkage of Pre-existing Flare Loops during a Subsequent Solar Eruption}

\begin{CJK*}{UTF8}{gbsn}
\correspondingauthor{Huadong Chen}
\email{hdchen@nao.cas.cn}

\author[0000-0001-6076-9370]{Huadong Chen (陈华东)}
\affiliation{National Astronomical Observatories, 
     Chinese Academy of Sciences, 
      Beijing 100101, People's Republic of China}
\affiliation{Key Laboratory of Solar Activity and Space Weather, National Space Science Center, Chinese Academy of Sciences, Beijing 100190, People's Republic of China}
\affiliation{School of Astronomy and Space Science, University of Chinese Academy of Sciences, Beijing 100049, People's Republic of China}

%\nocollaboration

\author[0000-0001-9315-7899]{Lyndsay Fletcher}
\affiliation{SUPA School of Physics and Astronomy, University of Glasgow, Glasgow, G12 8QQ, UK}
\affiliation{Oslo Rosseland Centre for Solar Physics University of Oslo, P.O.Box 1029 Blindern, NO-0315 Oslo,  Norway}

\author[0000-0001-8228-565X]{Guiping Zhou (周桂萍)}
\affiliation{National Astronomical Observatories, 
     Chinese Academy of Sciences, 
      Beijing 100101, People's Republic of China}
\affiliation{Key Laboratory of Solar Activity and Space Weather, National Space Science Center, Chinese Academy of Sciences, Beijing 100190, People's Republic of China}
\affiliation{School of Astronomy and Space Science, University of Chinese Academy of Sciences, Beijing 100049, People's Republic of China}

\author[0000-0003-2837-7136]{Xin Cheng (程鑫)}
\affiliation{School of Astronomy and Space Science, Nanjing University, Nanjing 210093, People's Republic of China}

\author[0000-0003-3699-4986]{Ya Wang (汪亚)}
\affiliation{Key Laboratory of Dark Matter and Space Astronomy, Purple Mountain Observatory, CAS, Nanjing, 210023, People's Republic of China}
%\affiliation{SUPA School of Physics and Astronomy, University of Glasgow, Glasgow, G12 8QQ, UK}

\author[0000-0002-9242-2643]{Sargam Mulay}
\affiliation{SUPA School of Physics and Astronomy, University of Glasgow, Glasgow, G12 8QQ, UK}

%\author[0000-0001-5685-1283]{Hugh Hudson}
%\affiliation{SUPA School of Physics and Astronomy, University of Glasgow, Glasgow, G12 8QQ, UK}

\author[0000-0002-2734-8969]{Ruisheng Zheng (郑瑞生)}
\affiliation{Institute of Space Sciences, Shandong University, Weihai 264209, People's Republic of China}

\author[0000-0002-5431-6065]{Suli Ma (马素丽)}
\affiliation{Key Laboratory of Solar Activity and Space Weather, National Space Science Center, Chinese Academy of Sciences, Beijing 100190, People's Republic of China}
\affiliation{National Astronomical Observatories, 
     Chinese Academy of Sciences, 
      Beijing 100101, People's Republic of China}

\author[0009-0005-2234-4319]{Xiaofan Zhang (张小凡)}
\affiliation{National Astronomical Observatories, 
     Chinese Academy of Sciences, 
      Beijing 100101, People's Republic of China}
\affiliation{Key Laboratory of Solar Activity and Space Weather, National Space Science Center, Chinese Academy of Sciences, Beijing 100190, People's Republic of China}
\affiliation{School of Astronomy and Space Science, University of Chinese Academy of Sciences, Beijing 100049, People's Republic of China}

\begin{abstract}
%Using the observations from the Atmospheric Imaging Assembly (AIA) on board the Solar Dynamics Observatory, 
We investigated two consecutive solar eruption events in the solar active region (AR) 12994 at the solar eastern limb on 2022 April 15. We found that the flare loops formed by the first eruption were involved in the second eruption. During the initial stage of the second flare, the middle part of these flare loops (E-loops) erupted outward along with the flux ropes below, while the parts of the flare loops (I-loops1 and I-loops2) on either side of the E-loops first rose and then contracted. Approximately 1 hour after the eruption, the heights of I-loops1 and I-loops2 decreased by 9 Mm and 45 Mm, respectively, compared to before the eruption. Their maximum descent velocities were 30 km s$^{-1}$ and 130 km s$^{-1}$, respectively. The differential emission measure (DEM) results indicate that the plasma above I-loops1 and I-loops2 began to be heated about 23 minutes and 44 minutes after the start of the second flare, respectively. Within $\sim$20 minutes, the plasma temperature in these regions increased from $\sim$3 MK to $\sim$6 MK. 
We proposed an adiabatic heating mechanism that magnetic energy would be converted into thermal and kinetic energy when the pre-stretched loops contract. Our calculations show that the magnetic energy required to heat the two high-temperature regions are 10$^{29}$-10$^{30}$ erg, which correspond to a loss of field strength of ∼2--3 G.
%Our observations, including the changes in the morphology and structure of the contracting magnetic loops, the appearance of a current sheet-like structure, and the heating of plasma overlying the contracting loops, suggest that magnetic reconnection may have occurred above the contracting coronal loops, which may be an important reason for the shrinkages of the coronal loops in addition to magnetic implosion.
%the brightenings at the top of the loops,
%within ~20-30 minutes
\end{abstract}


\section{Introduction} \label{sec:intro}
Flare loops are bright arch-like structures in the solar corona, appearing clearly in the late stage of solar flares \citep[e.g.,][]{fletcher11}.
 Their feet are rooted in the flare ribbons observed in the photosphere or chromosphere \citep[e.g.,][]{schmieder95, ding03, fletcher13, songq16, songy20}. 
%%In the standard CHSKP model,
The formation of flare loops is related to magnetic reconnection between opposite-polarity magnetic fields below the erupting magnetic structure and subsequent chromospheric evaporation \citep[e.g.,][]{lin00, fletcher11, liuw13, tian14}. 
%%Due to heat conduction and radiation cooling, flare loops generally appear in high-temperature wavebands first and then in the low-temperature wavebands in observations. 
During the emergence of a solar active region (AR), bright coronal loops can often be seen coming into being and overlying the AR at temperatures of millions or even tens of millions of kelvin in the corona \citep[e.g.,][]{lu24}. Since the plasma and magnetic field environment in the corona mostly meets force-free field conditions, it is generally believed that these coronal loops could reflect the local magnetic field structure \citep[e.g.,][]{wie12}. 

%that appear in the corona below erupting solar filaments or magnetic flux ropes (FRs) and above the photospheric magnetic neutral lines. 
%and gradual formation 
%bruzek64, 
%In space, the high-temperature flare loops are usually located at a higher position \citep[e.g.,][]{lin04}.
%In addition to the flare loops associated with eruptions,

%Flare loops are bright arch-like structures in the solar corona, appearing clearly in the late stage of solar flares \citep[e.g.,][]{fletcher11}.
%Their formation is believed to be related to magnetic reconnection and chromospheric evaporation \citep[e.g.,][]{lin00, fletcher13, liuw13, tian14}. 
Observations have indicated that both flare loops formed during flares and coronal loops near the eruption source regions often shrink \citep[e.g.,][]{sui03, ji07, liur09a, yan13, wangj16}. 
Typically, these shrinkages mainly occur in the early stages of flare loops' formation or the impulsive stages of eruptions \citep[e.g.,][]{sui04, liur09b}.
When combined with hard X-ray (HXR) observations, it is also possible to detect that the shrinkages of flare loops are often accompanied by the downward movements of HXR emission sources at the loop tops \citep[e.g.,][]{liy06, ji07}. 
Generally, compared to coronal loops, the shrinkages of flare loops are weaker, with a typical speed of $\sim$10 km s$^{-1}$ and a duration of several minutes; whereas the shrinkage speed of the coronal loops near eruptions can reach hundreds of km s$^{-1}$ and last for tens of minutes \citep[e.g.,][]{sui03, liur12b}. \citet{liuw13} reported shrinkage of flare loops that lasted about 2 hours, which they believed was evidence of reconnection outflows.
\citet{russell15} presented that the higher coronal loops oscillate while shrinking after the onset of a flare. In some cases, the coronal contractions during flares can even influence the structures and motions of photospheric sunspots \citep[e.g.,][]{bi16, xu17}.

%Before the widespread observation and reporting of coronal contractions, 
Theoretically, \citet{hudson00} proposed a hypothesis of ``magnetic implosion", in which the contraction of the corona during solar eruptions is considered to be caused by the reduction of magnetic pressure resulting from the release of magnetic energy. This proposal is supported by the calculations of \citet{janse07}, magnetohydrodynamic (MHD) simulations by \citet{wangj21} and is consistent with observations \citep[e.g.,][]{liur09b, liur12b, wangj18}. Additionally, the role of strong magnetic tension in the downward contractions of newly formed magnetic loops after reconnection appears to be important, as indicated by \citet{aschwanden04} and \citet{liuw13}. Furthermore, \citet{ji07} pointed out that the reduction of magnetic shear caused by the release of magnetic energy may also lead to the coronal magnetic arcades having smaller heights and spans. In the model of \citet{zuc17}, the vortices on both sides of erupting flux ropes (FRs) are used to explain the shrinking of the coronal loops around the eruption as described by \citet{dudik17}. However, it seems that it is hard for this model to explain why the contracting loops do not restore after the eruptions.
%\citep{liuw09}  
%(such as the flare loops)
%In the observations of \citet{liuw13}, it was found that the speeds of the contracting flare loops can reach up to 900 km s$^{-1}$, which is comparable to the expected coronal Alfv\'{e}n speed of 1000 km s$^{-1}$. 

%Usually, flare loops appear clearly in the late stage of solar flares. However, in some cases such as successive or homologous solar flares, coronal mass ejections(CMEs) \citep[e.g.,][]{zhang02, liul17, mitra20, zheng21} or when the magnetic FRs have stratified structures \citep[e.g.,][]{liur12a, kliem14, dhakal18}, the flare loops produced by the preceding flare can act as confining fields for the subsequent eruption. If the confining fields are not enough or if the driving forces below are too strong \citep[e.g.,][]{ji03, cheng11, chen13, chen14}, the flare loops may erupt in the subsequent eruption. 

So far, observations of flare loops' eruption have been rarely reported \citep[e.g.,][]{cheng10, joshi20}. As one of the important components of coronal mass ejections (CMEs), investigating their dynamic evolution and physical property changes is important for understanding the interactions between different structures of an erupting system. On 2022 April 15, two successive eruptions occurred in solar active region AR12994 at the eastern limb of the Sun from 00 UT to 12 UT. We analyzed the relationship between the two consecutive eruptions and studied the eruption and contraction of the pre-existing flare loops (formed by the first flare) during the second flare, as well as the changes in the physical characteristics of the surrounding plasma.
%chengx13

\section{Data and Observations} \label{sec:obser}
The Atmospheric Imaging Assembly \citep[AIA;][]{lemen12} on board the Solar Dynamics Observatory \citep[$SDO$;][]{pesnell12}, provides full-disk intensity images up to 0.5 R$_{\sun}$ above the solar limb with 0$\arcsec$.6 pixel size and 12 s cadence in seven EUV channels centered at 304 \AA\ (\ion{He}{2}, 0.05 MK), 171 \AA\ (\ion{Fe}{9}, 0.6 MK), 193 \AA\ (\ion{Fe}{12}, 1.3 MK and \ion{Fe}{24}, 20 MK), 211 \AA\ (\ion{Fe}{14}, 2 MK), 335\AA\ (\ion{Fe}{16}, 2.5 MK), 94 \AA\ (\ion{Fe}{18}, 7 MK), and 131 \AA\ (\ion{Fe}{8}, 0.4 MK and \ion{Fe}{21}, 11 MK), respectively \citep[e.g.,][]{odwyer10}. In our research, we used the AIA EUV data to illustrate the processes of the two consecutive eruption events and the relationship between them. Additionally, the AIA data in the 131 \AA, 171\AA, 193 \AA, 211 \AA, 335 \AA, and 94 \AA\ wavebands were utilized for differential emission measure (DEM) analysis \citep[][]{cheung15} on the source regions and nearby areas of the two eruptions. 
The DEM-weighted mean temperatures of these regions were calculated as 
$\bar{T}=\int DEM(T)\times TdT/\int DEM(T) dT$
over the range of log T [5.8, 7.2] \citep[e.g.,][]{wangy24}.
%\begin{equation}
%\end{equation}
%Assuming that the line-of-sight depth is equal to the width, we also estimated the densities of some areas.
Furthermore, we employed the coronagraph COR1 and COR2 data from the Sun Earth Connection Coronal and Heliospheric Investigation \citep[SECCHI;][]{howard08} onboard Solar Terrestrial Relations Observatory-A (STEREO-A) to demonstrate the structure of the CME caused by the first eruption.
%\citep[HMI;][]{schou12}
%The time-distance diagrams from the AIA 211 \AA\ and 131 \AA\ data aided us in analyzing the dynamic evolutions of the upward motions of the exploding coronal loops (E-loops) and associated FRs and the contraction motions of the imploding loops (I-loops1 and I-loops2) during the second eruption.
%, focusing on the current sheet formed by the first eruption and the EM and temperature changes in the plasma regions above I-loops1 and I-loops2 in the second eruption. 

\section{Results} \label{}
The two consecutive eruptions were separately associated with a C4.5 and M2.2 flare and resulted in a CME with a median velocity of 516 km s$^{-1}$ and 646 km s$^{-1}$ at 1.5--6 R$_{\sun}$, respectively (see the CACTus catalog, http://sidc.be/cactus). 
%The $GOES$-16 Soft X-ray (SXR) flux curves for the two flares can be found in Figure~5. 
Interestingly, we discovered that the flare loops created by the first eruption were involved in the second eruption. The observations in the multiple EUV wavebands of AIA clearly showed that the different parts of these flare loops underwent different dynamic processes. Some flare loops erupted to form the second CME, while other loops collapsed downward, demonstrating an implosive motion.

\subsection{The First Eruption} \label{subsec:}
In Fig.~1, the first eruption started around 00:00 UT on April 15. During the eruption, a hot twisted feature, proposed to be an erupting flux rope \citep[e.g.,][]{cheng11}, was observed to rise and erupt outward. 
%Its temperature was $\sim$4 MK according to the DEM results, as shown Figure~1(b1)--(b4).
%The DEM results indicate that the temperature of the FR was around 4 MK (see Figure~1(b1)). 
Subsequently, a narrow high-temperature structure gradually formed beneath the flux rope (see Fig.~1 (a2) and (b2)). 
It is very likely a current sheet formed by magnetic reconnection \citep[e.g.,][]{lil16, cheng18}.
Over the next 7 hours, the current sheet gradually widened, and its DEM-weighted mean temperature exhibited a rapid rise followed by a gradual decline (see Fig.~5(c1)). 
The box labeled as ``B1'' in Fig.~1(b3) denotes the area used for DEM analysis for the core of the current sheet. 
%Specifics regarding the DEM distribution, EM evolution, and temperature changes of this area are displayed in Figure~5(a1)-(c1). 
After about 06:00 UT, the flare loops began to emerge below the current sheet in the AIA EUV bands and continued to rise until the second eruption commenced at 10:30 UT.
The outlines of the flare loops were traced around 09:50 UT on the temperature map at the same time (see Fig.~1(b4)). Notably, just before the second eruption, there remained a high-temperature plasma region with a temperature of $\sim$4 MK above the flare loops ($\sim$2 MK). It might be related to the remnants of the reconnection. 
%on the verge of 
%Notably, just before the second eruption, there remained a distinct high-temperature plasma region above the flare loops, with a temperature of approximately 4MK, while the loops themselves maintained a temperature of around 2MK.
%, which themselves had a temperature of about 2 MK. 
%This eruption was a typical magnetic FR eruption event. 

%This eruption resulted in a CME traveling at a median velocity of $\sim$520 km/s. 
In Fig.~1(c), we overlaid the simultaneous SECCHI COR2 (red), COR1 (green) data, and AIA 131 \AA\ (cyan) image at around 03:54 UT to show the associated CME.
%The rectangular box in the composite image indicates the field of view (FOV) of Figure~1(a1)-(b4). 
The observation of COR2 clearly reveals the three-part structure of the CME: the erupting front, the dark cavity, and the bright core \citep[e.g.,][]{songh23}. In the COR1 data, a bright narrow structure (indicated by the arrow) appears to connect the bright core, i.e. the erupting flux rope, and the hot current sheet on the AIA 131 \AA\ image. 
It is worth noting that similar observations have been documented previously, and these narrow structures, with a length of about several solar radii, are considered part of the entire current sheet structure \citep[e.g.,][]{lin05}. The DEM results indicate that the high-temperature regions in the current sheet are primarily concentrated close to the sun's surface, consistent with other observations \citep[e.g.,][]{liuw13}.
% \citet{liuw13} and \citet{lil16}.

\subsection{The Subsequent Eruption} \label{subsec:}
At 10:30 UT, a second eruption took place from the same active region. Based on AIA observations, it was discovered that the flare loops that formed in the first eruption participated in the subsequent eruption, but not all of them erupted outward. Fig.~2 primarily illustrates the different motions of these loops in the second eruption. The exploding loops (E-loops) mainly constituted the middle section of the entire flare loops. On the other hand, the loops on the northern and southern sides initially ascended with the eruption and then rapidly contracted downward or imploded (I-loops1 and I-loops2). In Fig.~2(a1), the yellow and blue arrows approximately indicate the directions of motion for E-loops, I-loops1, and I-loops2, respectively. The intensity images at AIA 171 \AA\ in Fig.~2(b1) and (b2) distinctly show some fine structures of the coronal loops.
%that erupted outward 
 %in the relatively cooler temperature band($\sim$0.8 MK). 
In the high-temperature waveband at AIA 131 \AA\ ($\sim$11 MK), as the E-loops inflated, hot magnetic flux ropes gradually rose from below and erupted outward, ultimately resulting in the formation of the second CME.
%as observed by the coronagraph. 
In Fig.~2(c), the composite double-colour image composed of the 211 \AA\ image at 11:12:21 UT and the 131 \AA\ image at 11:12:18 UT presents the spatial relationship between the flux ropes and the E-loops, I-loops1, and I-loops2.
We also created composite double-colour images using the 211 \AA\ and 94 \AA\ images to study the variations in emissions from the high-temperature plasma ($\sim$7 MK) surrounding the cooler 211 \AA\ loops ($\sim$2 MK).
This is shown in Fig.~2 (a3) as one example.
%we found that these loops with a height of about 80 Mm did not all move outward. 
%The median velocity of the second CME is about 650 km s$^{-1}$. 

\subsubsection{Dynamics of E-loops and Flux Ropes' Eruptions} \label{subsec:}
According to the positions of the three slits S1, S2, and S3 in Fig.~2(a3) and (c), and utilizing composite double-colour images from the AIA in 211 \AA\ and 131 \AA, and 211 \AA\ and 94 \AA, we generated time-distance maps (Fig.~3(a)--(c)), displaying the changes in height of E-loops, flux ropes, I-loops1, and I-loops2 during the second eruption. In Fig.~3(a)--(c), the dotted curves represent the positions of the tops of E-loops, flux ropes, I-loops1, and I-loops2. Observations at AIA 131 \AA\ indicate the presence of two erupting magnetic flux ropes or two distinct parts of one flux rope structure \citep[FR1 and FR2; e.g.,][]{liur12a, kliem14, zheng21} beneath the E-loops, as they exhibit different dynamic evolutions.
Additionally, Fig.~3(b) and (c) indicate that the plasma emissions above I-loops1 and I-loops2 in the 94 \AA\ channel were significantly enhanced after the loops' shrinkages.

Fig.~3(d)-(e) presents the time-velocity and time-acceleration curves for E-loops, flux ropes FR1 and FR2. The velocities and accelerations of E-loops and flux rope FR1 are quite similar in both value and evolution, indicating that the rise of flux rope FR1 caused the expansion of overlying E-loops.
%the rise of FR1 caused the expansion of overlying E-loops.
Notably, the velocities of E-loops and flux rope FR1 do not increase uniformly. 
Around 10:51 UT (indicated by the blue vertical line), the velocity of flux rope FR1 began to decrease, and about 5 minutes later, at 10:56 UT (indicated by the red vertical line), the velocity of E-loops also started to decline. 
%As shown in Figure~3(a), the slowdowns in the height increase of FR1 and E-loops (indicated by the arrows) also confirm this trend. 
Concurrently, below FR1, the flux rope FR2 began to ascend and erupt outward. After 11:20 UT, its acceleration increased rapidly from tens of m s$^{-2}$ to more than 200 m s$^{-2}$. Potentially due to the impact of the erupting flux rope FR2, around 11:00 UT (indicated by the pink vertical line), the velocities of E-loops and flux rope FR1 began to increase again. These changes in the motion patterns of E-loops and flux ropes reflect the complexity of solar eruption regarding the competition and interaction between the driving forces and magnetic confinements \citep[e.g.,][]{ji03, shen11, chen13, chen14, chen15, jiang16}.

%The velocities of E-loops, FR1, and FR2 started at tens of km s$^{-1}$ and gradually increased to $\sim$200 km s$^{-1}$ within $\sim$1 hour from the eruption onset. However, the associated CME had a median velocity of $\sim$650 km s$^{-1}$, suggesting that the eruption system continued to accelerate after 11:30 UT. It is possible that magnetic reconnection below the erupting FR played a significant role in the acceleration process \citep[e.g.,][]{lin00, cheng18}.

%the observations from the LASCO coronagraph revealed that 
% at 10:30 UT.
%were initially not very high
%During the process of magnetic FR or filament eruption or CME formation, especially in the early stages, various factors associated with the eruption driving forces and magnetic confinements are competing and interacting with each other. The result of this competition and interaction will determine the success or failure of the eruption and whether a CME can be formed \citep[e.g.,][]{ji03, shen11, chen14, chen15, jiang16}.
%and exploded together with FR2. 

\subsubsection{Dynamics of I-loops1 and I-loops2's Shrinkages} \label{subsec:}
%During the second flare, the AIA observations reveal not only the eruption of certain flare loops (E-loops) but also the simultaneous contractions of the flare loops (I-loops1 and I-loops2) on both sides of the E-loops. 
In Fig.~3(b) and (c), the time-height maps separately indicate the height changes of I-loops1 and I-loops2 as the E-loops and flux ropes erupted outward. 
%show time-height maps derived from the AIA 211 \AA\ images along the slits S2 and S3, indicating changes in the height 
Their velocity-time and acceleration-time curves are depicted in Fig.~3(f) and (g), respectively. It is evident that before shrinking both I-loops1 and I-loops2 first rose, likely due to compression from the eruptions of E-loops and flux ropes. 
In the subsequent descent phase, I-loops1 contracted before I-loops2, but its descent was less violent than that of I-loops2. Our results indicate that the height of I-loops1 dropped by about 9 Mm compared to before the eruption, with a maximum descent speed of $\sim$30 km s$^{-1}$ and a maximum descent acceleration of $\sim$50 m s$^{-2}$. On the other hand, I-loops2 underwent two more violent contractions, with a height drop of about 45 Mm compared to before the eruption, a maximum descent speed of $\sim$130 km s$^{-1}$, and a maximum descent acceleration of $\sim$1 km s$^{-2}$. The contraction dynamics of I-loops1 and I-loops2 are similar to those of the coronal loops near eruption source regions as observed previously \citep[e.g.,][]{yan13, wangj16}.

%A notable feature of this study is that o
%Additionally, we observed that the intensities of the tops of these shrinking loops in various AIA EUV wavebands are significantly enhanced. For example, we highlighted the brightenings at 211 Å with the arrows in Figure 3(b) and (c).  These brightenings are probably caused by the accumulation of material after the loops shrank.

\subsubsection{The Heated Plasma regions above I-loops1 and I-loops2} \label{subsec:}
Another interesting aspect of this work is that we found that the plasma above I-loops1 and I-loops2 were heated, as demonstrated by the appearance of the AIA 94 \AA\ brightenings in the double-colour images (see Fig.~3b and c).
The temperature map in Fig. 4 illustrates the DEM-weighted mean temperature distribution of the eruption source region and the surrounding area from the eruption onset (10:30 UT) to about one hour later (11:32 UT) when I-loops1 and I-loops2 stopped shrinking. 
%We outlined the E-loops, I-loops1, and I-loops2 on the temperature maps, which are more visible on the 211 \AA\ images. This allowed for a better understanding and analysis of the temperature conditions of these loops and their surrounding areas. 
% causing their temperatures to increase
Similar to Fig.~1(b4), Fig.~4(a) demonstrates that before the second eruption, there was a relatively high-temperature plasma region above the E-loops, I-loops1, and I-loops2, i.e. the flare loops formed by the first flare. As the E-loops and flux ropes expanded and erupted, this high-temperature area initially faded gradually with its middle and southern parts even disappearing after 10:56 UT, and then, around 10:54 UT, its northern part began to strengthen and expand (indicated by the arrow in Fig.~4(b)). 
At around 11:12 UT, this enhanced high-temperature area connected with the contracting I-loops1 and continued to expand until around 11:35 UT. Until around 11:15 UT, on the southern flank of the erupting E-loops and flux ropes, a new high-temperature area began to emerge above I-loops2  (indicated by the arrow in Fig.~4(d)), which also continued to develop and expand until around 11:35 UT.

\subsection{DEM Analyses and Energy Estimations} \label{subsec:}
We conducted DEM analyses on the cores of the high-temperature plasma regions above I-loops1 and I-loops2, as indicated by the boxes ``B2'' and ``B3'' in Fig. 4(f). 
The black curves in Fig.~5(b1)--(b3) and (c1)--(c3) display the DEM distributions, EM evolutions, and DEM-weighted mean temperature changes of the two regions, respectively. 
For comparison, the DEM analysis results of area B1, i.e. the core region of the current sheet formed in the first eruption, are shown in Fig.~5(a1)--(a3).
%we also performed DEM analysis on the core region of the current sheet structure formed by the first eruption (B1 in Figure~1(b3)). 
%The times of the DEM distributions of B1, B2, and B3 are separately indicated by the vertical lines in their time-EM and time-temperature curves diagrams. 
The DEM distributions of B1, B2, and B3 all exhibit a bimodal structure, indicating a multi-temperature composition and structure of the plasma within them. The temperature corresponding to the low-temperature peak is about 1.6 MK (log T$\approx$6.2), while the temperature corresponding to their high-temperature peak is about 6.3 MK (log T$\approx$6.8), which approximately corresponds to the peak temperature of the AIA 94 \AA\ band. Using the method provided by \citet{del13}, we found that the main contributions ($\sim$98\%) to the 94 \AA\ band are from the \ion{Fe}{18} emission.
Additionally, it can be seen that the DEM results are not well constrained
%have rather large uncertainties 
in the ranges where log T is less than 5.8 and greater than 7.2. Consequently, we excluded the DEM distribution in these ranges when calculating the temperatures.

%To analyze the relationships between the EM and temperature changes of the plasma in the three regions and the related flares, 
We also plotted the Geostationary Operational Environment Satellite (GOES-16) soft X-ray (SXR) 1--8 \AA\ fluxes of the two flares in the middle and bottom panels of Fig.~5. For the current sheet formed in the first flare, represented by area B1, both the EM and temperature began to rise at the same time as the SXR flux of the flare. However, the peak of the temperature and EM occurred earlier and later, respectively, than that of the flare's SXR flux. Overall, region B1's temperature gradually decreased after rapidly reaching a peak of $\sim$10 MK, and its EM rising phase experienced a longer time, which is similar to previous observations \citep[e.g.,][]{sun14, lil16}.
For the high-temperature areas B2 and B3, their EM and temperature began to rise about half an hour later than the start of the second flare (see Fig. 5 panels b2--c2 and b3--c3). We noticed that this flare's SXR flux did not decrease immediately after reaching a peak at around 11:00 UT, but had a new peak $\sim$50 minutes later. The EM and temperature rises of B2 and B3 seem to correspond well to the appearance of the second SXR flux peak of the flare. The peak temperature of the plasma in B2 and B3 is about 6 MK. Additionally, the change trends of EM and temperature in B2 are relatively synchronized, which is significantly different from B1, reflecting that the mode or physical process of energy release and heating in the two regions may be different. 
The EM and temperature evolution of B3 is similar to that of B2, but after 11:30 UT, due to the influence of the second flare, these two high-temperature areas are mixed with the surrounding high-temperature regions and are hard to identify.

Based on the assumption that the line-of-sight depths ($l$) of the interested regions equal to their projected widths, the electron number density were estimated by $n_e= \sqrt{EM/l}$.
Then, the total thermal energy is approximately $E_t=(3/2) 2n_ek \bar{T} V$ in the case of full-ionization. Here, $k$ is the Boltzmann constant and $V$ is the volume of the entire interested region. 
The dark green and blue curves in Fig. 5 display the evolutions of $n_e$ and $E_t$ for the three regions, respectively. 
Overall, the $n_e$ of B1 and B3 varies in the range of $\sim$[1.5, 2.5] $\times$ 10$^{9}$ cm$^{-3}$, while the $n_e$ of B2 are relatively smaller with a mean value of $\sim$1.0 $\times$ 10$^{9}$ cm$^{-3}$, likely due to its higher altitude.
For $E_t$, its change in the current sheet of the first flare is well synchronized with the EM evolution of B1; while the thermal energy evolutions of the two high-temperature regions in the second flare are more consistent with their temperature changes. The peak values of $E_t$ for the three regions are 6.9, 12.6, and 3.4 $\times$ 10$^{29}$ erg, respectively. 
Using the shrinkage velocities $v_s$ of I-loops1 and I-loops2, we can also estimate the kinetic energies of the plasma in the two high-temperature regions as 
$E_k=(1/2)n_e m_h v_s^2 V$, in which $m_h$ is the proton mass.
Taking $v_s$ as 30 and 130 km s$^{-1}$, the peak values of $E_k$ can be separately derived as 0.4 and 2.0 $\times$ 10$^{28}$ erg, which are much less than those of $E_t$.
If considering an adiabatic heating mechanism due the shrinkages of I-loops1 and I-loops2 and neglecting the other energies, the magnetic energies $E_m$ converted into the thermal and kinetic energies should be the sums of $E_t$ and $E_k$. According to our calculations, the peak values of $E_m$ are 12.7 and 3.6 $\times$ 10$^{29}$ erg for I-loops1 and I-loops2, respectively.
%, which correspond to a loss of magnetic field strength of $\sim$2--3 G.

%above the contracting loops I-loops1 and I-loops2, 
%in the following two hours

\section{Summary and Discussion} \label{sec:summary}
In this work, we report two consecutive solar eruptions in the same solar AR, both linked to the eruptions of hot magnetic flux ropes. The two eruptions produced a C4.5 and an M2.2 flare, along with CMEs traveling at speeds of $\sim$500 km s$^{-1}$ and $\sim$650 km s$^{-1}$, respectively. It has been found that during the second eruption, the pre-existing flare loops created by the first eruption partially erupted and shrank. 
In general, the shrinkages of coronal loops around the source region of an eruption are not uncommon \citep[e.g.,][]{liur09a, yan13}, but it is unusual to observe both the partial eruption and the simultaneous contraction of the loops during the same event, especially when these coronal loops are part of the flare loop system formed by a previous eruption. The eruption of the flare loops are obviously driven by the erupting flux ropes below. 

Sometimes, solar eruptions occur without any obvious CMEs being detected \citep[e.g.,][]{green02, zhou03, ma10}. These events are referred to as failed eruptions \citep[e.g.,][]{ji03, alex06, liuy09} or confined flare activities \citep[e.g.,][]{wangy07, guo10, chen15, lit19}. Both theoretical \citep[e.g.,][]{torok05, fan07, guo10, jiang12} and observational \citep[e.g.,][]{shen11, zheng12, chen13, chen15} studies have indicated that the strong confinement from the overlying magnetic field is the primary cause of these failed eruptions. 
The occurrence of magnetic loop shrinkage or even flux rope contraction in such failed eruptions remains uncertain, as no relevant observational evidence has been found so far.
\citet{zhang17} suggest that complex flare loop structures in the well-known solar active region AR12192 \citep[e.g.,][]{chen15, sun15, thal15} may be responsible for confined flares. However, it is unclear whether flux ropes and overlying magnetic loops contract during these confined flare.

In a fully confined flare, the magnetic energy is converted into thermal energy and kinetic energy of accelerated particles, which means that energy has been taken from the magnetic field \citep[e.g.,][]{fletcher11, priest14}. So there would be a corresponding decrease in the magnetic pressure, with energy overall being conserved in the volume of the confined flare. 
However, the system is not closed, and over time the thermal energy, and also the kinetic energy of accelerated particles, is ultimately radiated away and lost. 
So although the gas pressure is initially high due to the increase of thermal energy, over some timescale the gas pressure will also decrease and it would be therefore expected that there would be contraction of the overlying loops. It might not be so dramatic as in the case of an escaping flux rope removing a lot of magnetic energy from the system.
Observations by \citet{chen13} showed that during an X1.9-class confined flare associated with a filament eruption, the overlying magnetic arches halted the filament eruption and did not immediately collapse downward, but instead rose slowly for a period. 
In our observations, the early motion of the imploding loops I-loops1 and I-loops2 is similar to that in a confined flare, initially rising with the erupting flux rope and then halting. However, the subsequent and rapid contraction of these loops may be linked to the successful eruption of the E-loops and the flux ropes, as their escape results in a significant decrease in magnetic pressure across the entire flare area \citep{hudson00}.

%a failed flux rope eruption does not cause a significant decrease in magnetic pressure. However, a flare resulting from magnetic energy release beneath the confining field leads to an increase in thermal energy and gas pressure. Consequently, it is challenging to anticipate the contraction of overlying loops in such a scenario. 
%The shrinkages of other parts of them are likely due to magnetic implosion caused by the reduction of magnetic pressure \citep{hudson00}. 
%Would the shrinkage of magnetic loops or even flux ropes occur in such failed eruptions? As far as we know, no relevant observational evidence has been found so far. 

Interestingly, two high-temperature regions separately appeared above the contracting flare loops in our observations.
Here, We propose an adiabatic heating mechanism. In this model, stretching the field corresponds to increasing the magnetic energy, which can then be converted to thermal and kinetic energies when contraction happens. This scenario is somewhat similar to electron induction acceleration in a betatron.
In our observations, the loops had previously expanded, driven upwards by the erupting flux rope, as seen in Fig. 3b and 3c, and therefore energised. We estimate that the magnetic energies required to heat the two high-temperature regions are 10$^{29}$-10$^{30}$ erg, which correspond to a loss of field strength of ∼2--3 G. If the field strengths of the two regions are estimated as $\sim$10 G \citep{dulk78}, that means that about 20--30$\%$ of the total magnetic energy would be converted to heat the plasma during the contraction.
It should be noted that this decrease in magnetic field strength is discussed assuming that the volumes of the entire interested regions remain unchanged. If only considering the volumes of the shrinking loops, their field strengths should increase accordingly. Assuming that the magnetic moments are conserved, i.e. $\mu=E_{\bot}/B=k\bar{T}/B=C$, and taking $B_0$, $\bar{T_0}$, and $\bar{T_1}$ as 10 G, 3 MK, and 6 MK, then the field strength of the loops will increase to 20 G at the time of the peak temperature.
%$(1/2)m_hv_{\bot}^2/B=k\bar{T}/B=constant$

When the heated plasma regions began to appear between 10:55 UT and 11:15 UT, the velocities of E-loops and flux ropes are rather small with the values of $\sim$100 km s$^{-1}$, which are less than the local Alfv\'{e}n speed of $\sim$140--220 km s$^{-1}$ (from $v_a=B/\sqrt{4\pi \rho}$, where $\rho$ is the mass density). It is hard to believe that the eruption disturbances would lead to the production of coronal shocks \citep[e.g.,][]{ma11} and the consequent shock heatings.
On the other hand, if magnetic reconnection could occurred above the shrinking loops, for instance by squeezing the opposite-polarity magnetic fields above I-loops1 and I-loops2 and/or due to the inflows caused by the reduction of gas and magnetic pressure during the E-loops and flux ropes' eruption, then the newly-formed loops may also contract and compress I-loops1 and I-loops2 owing to magnetic tension. Additionally,  magnetic reconnection can explain the appearance of the two high-temperature regions above the shrinking loops. 
However, due to the projection effect, more evidence of magnetic reconnection still needs to be found for this or other similar events in the future.

\begin{acknowledgments}
%We thank Feng Chen in Nanjing University and Baolin Tan, Shuhong Yang, Leping Li, Chunlan Jin, and Yijun Hou in National Astronomical Observatories, Chinese Academy of Sciences for the helpful discussions. 
The $SDO$ data are courtesy of NASA, the $SDO$/AIA, and $SDO$/HMI science teams. The STEREO/SECCHI data are produced by an international consortium: NRL, LMSAL, NASA, GSFC (USA); RAL (UK); MPS (Germany); CSL (Belgium); and IOTA, IAS (France). This work is supported by the National Key R\&D Program of China 2021YFA1600502, 2021YFA1600503 (2021YFA1600500), 2022YFF0503800, and 2022YFF0503003 (2022YFF0503000), the Strategic Priority Research Program of Chinese Academy of Sciences (Grant No. XDB 41000000), the Key Research Program of Frontier Sciences, CAS (Grant No. ZDBS-LY- SLH013), Beijing Natural Science Foundation (1202022), the National Natural Science Foundations of China (NSFC; 12073042, 12350004, and 12273061) and Yunnan Academician Workstation of Wang Jingxiu (No. 202005AF150025). L.F. acknowledges support from grants ST/L000741/1 and  ST/X000990/1 made by UK Research and Innovation's Science and Technology Facilities Council (UKRI/STFC). H.D.C. was also supported by the Chinese Academy of Sciences (CAS) Scholarship.
\end{acknowledgments}

%% The reference list follows the main body and any appendices.
%% Use LaTeX's thebibliography environment to mark up your reference list.
%% Note \begin{thebibliography} is followed by an empty set of
%% curly braces.  If you forget this, LaTeX will generate the error
%% "Perhaps a missing \item?".
%%
%% thebibliography produces citations in the text using \bibitem-\cite
%% cross-referencing. Each reference is preceded by a
%% \bibitem command that defines in curly braces the KEY that corresponds
%% to the KEY in the \cite commands (see the first section above).
%% Make sure that you provide a unique KEY for every \bibitem or else the
%% paper will not LaTeX. The square brackets should contain
%% the citation text that LaTeX will insert in
%% place of the \cite commands.

%% We have used macros to produce journal name abbreviations.
%% \aastex provides a number of these for the more frequently-cited journals.
%% See the Author Guide for a list of them.

%% Note that the style of the \bibitem labels (in []) is slightly
%% different from previous examples.  The natbib system solves a host
%% of citation expression problems, but it is necessary to clearly
%% delimit the year from the author name used in the citation.
%% See the natbib documentation for more details and options.
%refer12

%\bibliography{sample631}{}
%\bibliographystyle{aasjournal}

\begin{figure}
\epsscale{1}
\plotone{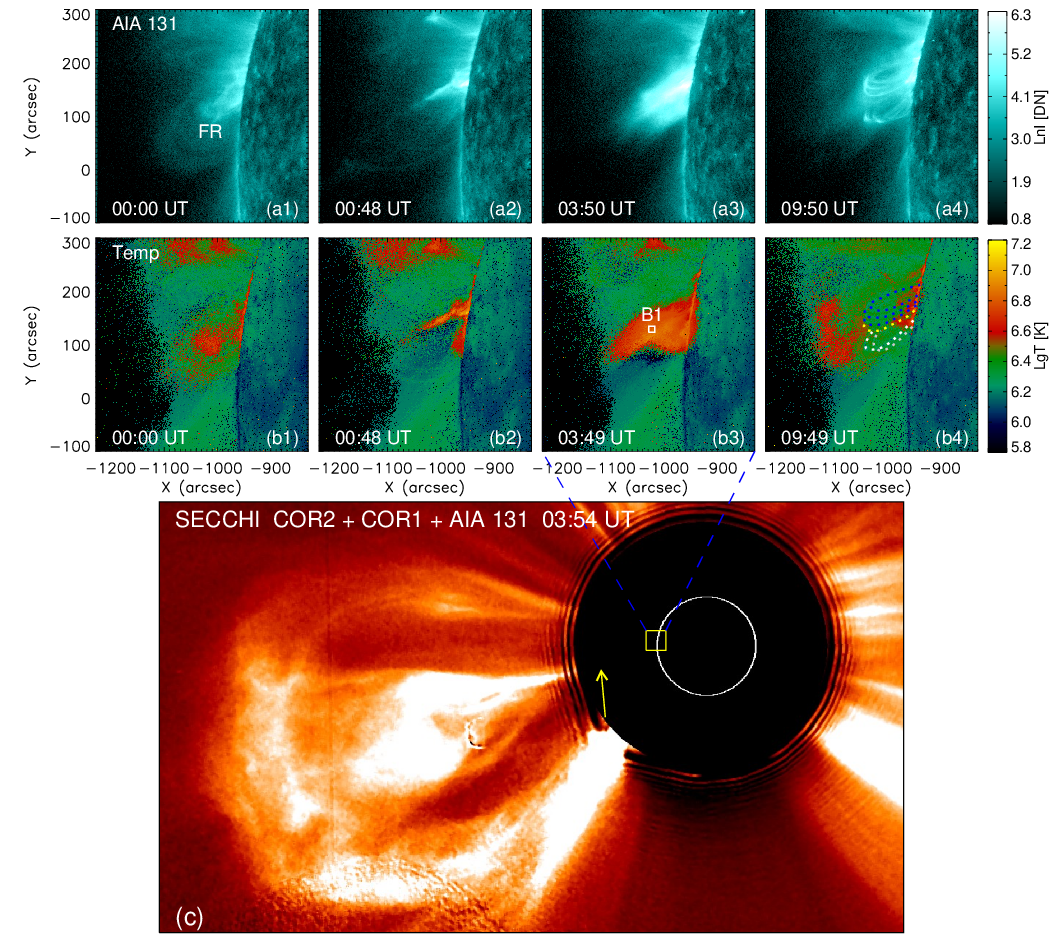}
\caption{(a1)--(a4) The AIA 131 \AA\ images display the hot erupting flux rope (FR), current sheet, and flare loops during and after the first eruption; (b1)--(b4) The corresponding DEM-weighted mean temperature maps; (c) The SECCHI COR2 (red) white-light data overlaid with the SECCHI COR1 (green) white-light and AIA 131 \AA\ (cyan) image. 
For more details, see Movie~1.
%In panel (b3), area B1 marks where we calculated the difference emission measure (DEM), emission measure (EM), and temperature of the current sheet. In panel (b4), the dotted curves display profiles of the flare loops. %The box in panel (c) shows the field of views (FOVs) of the AIA 131 \AA\ images and temperature maps, while the arrow points to a long and narrow structure in the SECCHI COR1 data.
\label{fig1}}
\end{figure}

\clearpage
%\clearpage
%\begin{figure}[ht!]
\begin{figure}
\epsscale{1}
\plotone{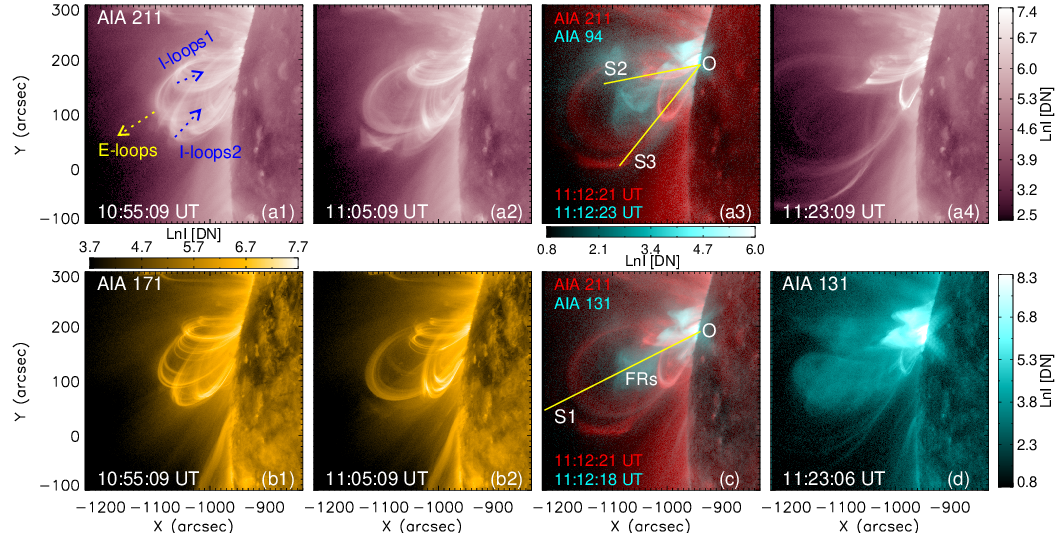}
\caption{(a1)--(a4) The AIA 211 \AA\ images and composite double-colour image consisting of the 211 \AA\ image (in red) and the 94 \AA\ image (in cyan) show the exploding loops (E-loops) and imploding loops (I-loops1 and I-loops 2) during the second eruption. Panels (b1) to (b2) display the E-loops and I-loops in the AIA 171 \AA\ channel. (c) The composite double-colour image consisting of the 211 \AA\ image (in red) and the 131 \AA\ image (in cyan). Panel (d) shows the hot erupting flux rope. 
%Yellow and blue arrows roughly indicate the erupting and contracting directions of the E-loops and I-loops, respectively. 
The slits S1, S2, and S3 indicate the positions from which we obtain the time-distance maps shown in Fig.~3. 
For more details, see Movie~2.
\label{fig2}}
\end{figure}

\begin{figure}
\epsscale{1}
\plotone{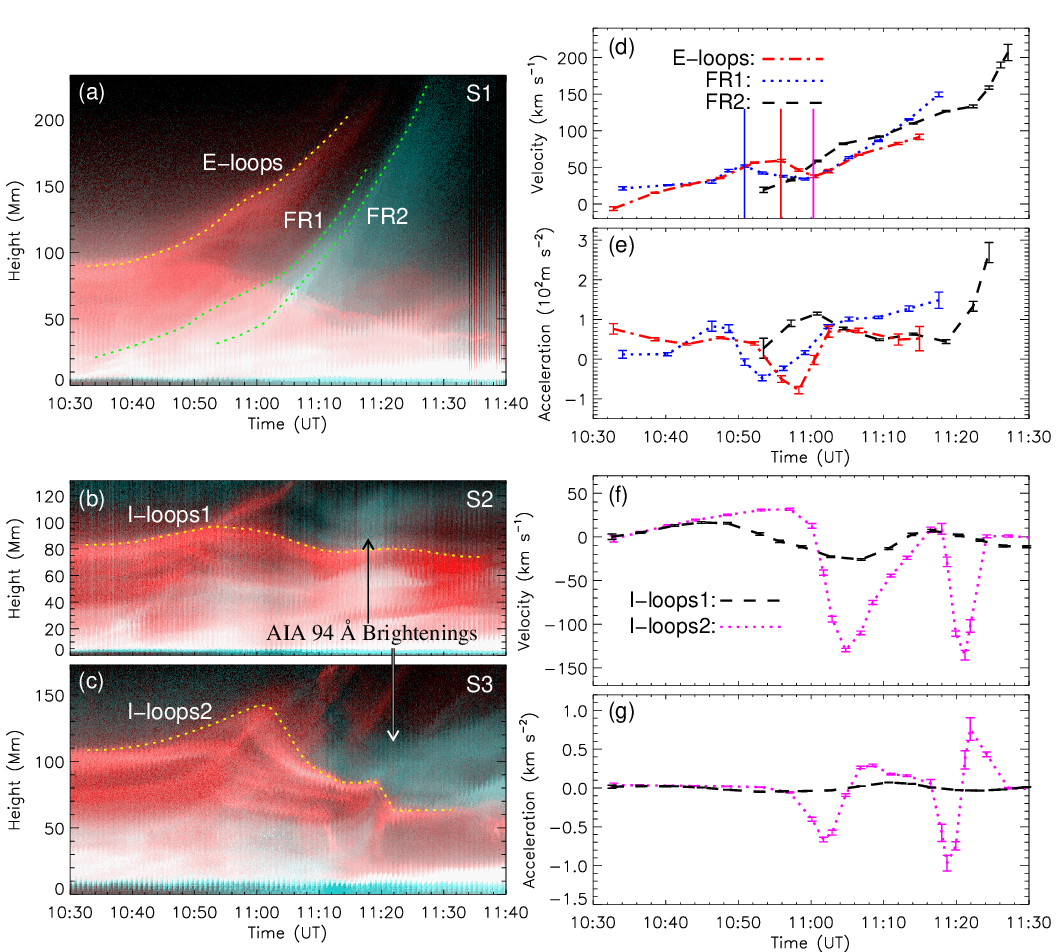}
\caption{(a)--(c) The time-distance maps are from the slits S1, S2, and S3, respectively. The dotted curves in panels (a)--(c) display the evolutions of the tops of the E-loops, flux ropes FR1 and FR2, I-loops1, and I-loops2. (d)--(e) The time-velocity and time-acceleration curves of the E-loops, flux ropes FR1 and FR2, respectively. (f)--(g) Similar to (d)--(e), but for I-loops1 and I-loops2. 
%The blue and red vertical lines in panel (d) indicate the times when the velocities of FR1 and E-loops began to decrease separately. The purple vertical line in panel (d) shows the time at which the velocities of FR1 and E-loops began to increase again.
%The time-velocity and time-acceleration curves of the I-loops1 and I-loops2, respectively.
%The arrows in panels (b) and (c) indicate the AIA 94 \AA\ brightenings above I-loops1 and I-loops2 after contractions.
\label{fig3}}
\end{figure}

\begin{figure}
\epsscale{1}
\plotone{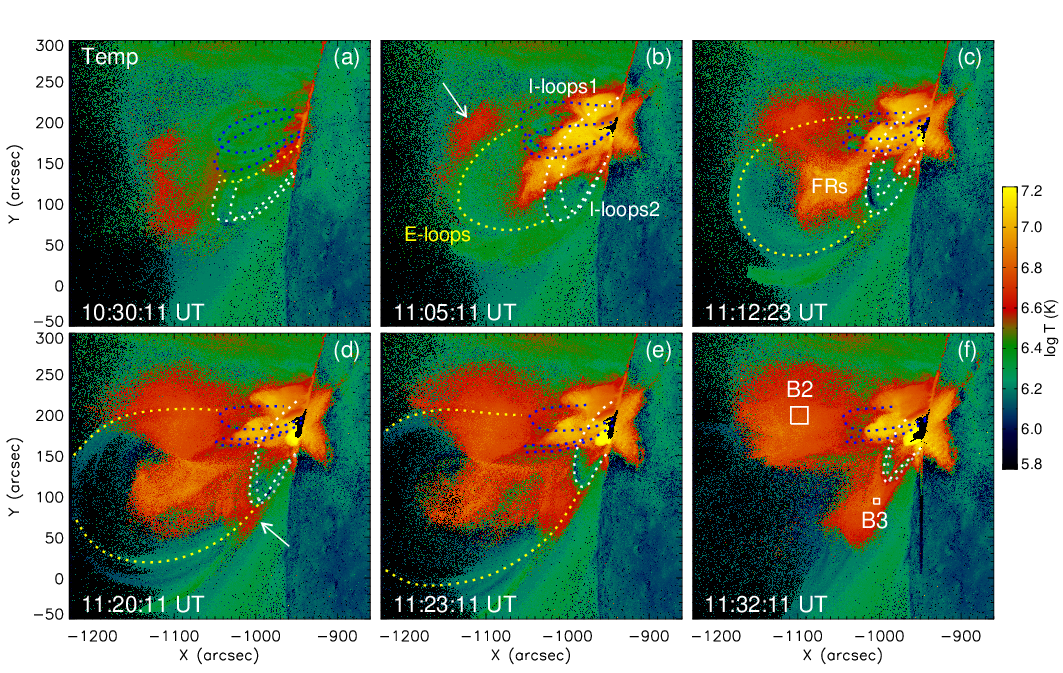}
\caption{(a)--(f) The DEM-weighted mean temperature maps of the source region during the second eruption. The yellow, blue, and white or grey dotted curves approximately describe the profiles of the E-loops, I-loops1, and I-loops2, respectively. The arrows in panel (b) and panel (d) point to the heated plasma regions above I-loops1 and I-loops2, respectively. 
%The boxes ``B2'' and ``B3'' in panel (f) indicate the areas where we calculated the DEMs, EMs, and temperatures of the two heated plasma regions.
\label{fig4}}
\end{figure}
%\clearpage

\begin{figure}
\epsscale{1}
\plotone{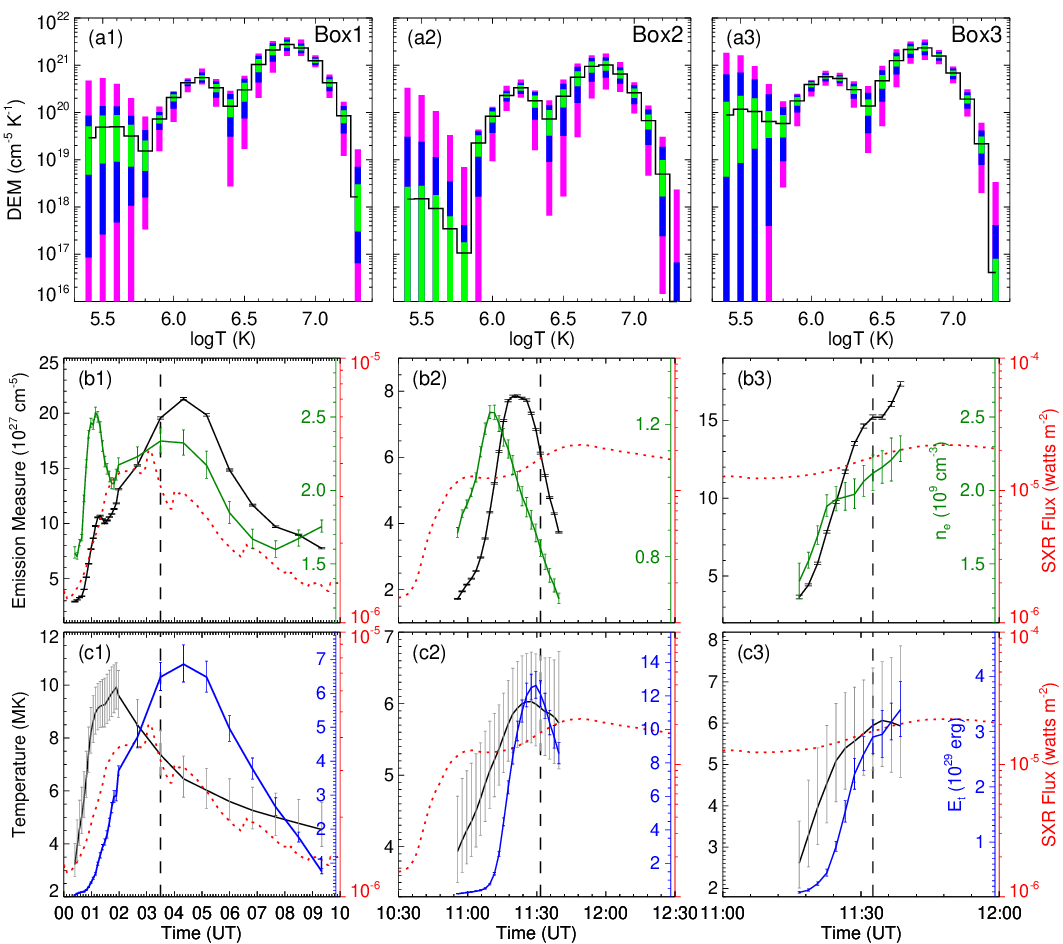}
\caption{(a1)--(c1) The DEM and the time profiles of EM, electron number density ($n_e$), and DEM-weighted mean temperature with 1$\sigma$ uncertainties for the area B1. 
The blue curve is the time profile of thermal energy ($E_t$) with 1$\sigma$ errors for the entire current sheet in the first flare.
(a2)--(c2) and (a3)--(c3) are similar to (a1)--(c1), but for ``B2'' and ``B3'', respectively. The black curve in panels (a1)--(a3) represents the best-fit DEM distribution. The green, blue, and pink rectangles in (a1)--(a3) separately denote the regions containing 50\%, 51--80\%, and 81--95\%\ of the Monte Carlo solutions. The vertical dashed lines in panels (b1)--(b3) and (c1)--(c3) indicate the times at which the DEMs in panels (a1)--(a3) are, respectively. The red dotted curves in the middle and bottom panels represent the $GOES$-16 Soft X-ray 1--8 \AA\ fluxes for the two flares.
%The blue curves in panels (c1)--(c3) indicate the thermal energy ($E_t$) changes of the entire current sheet in the first flare and the northern and southern high-temperature regions in the second flare. 
%from the  channel
%with their 1$\sigma$ error bars
\label{fig5}}
\end{figure}
\end{CJK*}

\end{document}